\documentclass[preprint,12pt]{elsarticle}

\usepackage{amsmath}
\usepackage{amssymb}
\usepackage{graphicx}
\usepackage{dcolumn}
\usepackage{bm}
\usepackage{xcolor} 
\usepackage{textcomp} 
\usepackage{verbatim}
\usepackage{textgreek}
\newcommand{\Intd}{\mathrm{d}}
\newcommand{\kBT}{k_\mathrm{B} T}
\newcommand{\muFS}{\mu^\mathrm{FS}}
\newcommand{\muNS}{\mu^\mathrm{NS}}
\journal{J. Coll. Int. Sci.}

\begin{document}

\begin{frontmatter}

\title{Microparticle Brownian Motion near an Air-Water Interface Governed by Direction-Dependent Boundary Conditions}

\author{Stefano Villa\footnote{Max Planck Institute for Dynamics and Self-Organization, G\unexpanded{ö}ttingen (Germany)}, Christophe Blanc\footnote{Laboratoire Charles Coulomb (L2C), UMR 5221 CNRS-Universit\unexpanded{é} de Montpellier (France)}, Abdallah Daddi-Moussa-Ider\textsuperscript{1}, Antonio Stocco\footnote{Institut Charles Sadron, CNRS UPR22, University of Strasbourg (France)}, Maurizio Nobili\textsuperscript{2,*}}

\date{\textsuperscript{*}\textit{maurizio.nobili@umontpellier.fr}}

%\author[label1]{Stefano Villa}
%\author[label2]{Christophe Blanc}
%\author[label1]{Abdallah Daddi-Moussa-Ider}
%\author[label3]{Antonio Stocco}
%\author[label2]{Maurizio Nobili\corref{cor1}}

%\cortext[cor1]{Corresponding author: \textit{maurizio.nobili@umontpellier.fr}}
%\affiliation[label1]{organization={Max Planck Institute for Dynamics and Self-Organization},
%             %addressline={},
%             city={G\unexpanded{ö}ttingen},
%             postcode={37077},
%             %state={},
%             country={Germany}}
             
%\affiliation[label2]{organization={Laboratoire Charles Coulomb (L2C), UMR 5221 CNRS-Universit\unexpanded{é} de Montpellier},
%             %addressline={},
%             city={Montpellier},
%             %postcode={},
%             %state={},
%             country={France}}
             
%\affiliation[label3]{organization={Institut Charles Sadron, CNRS UPR22, University of Strasbourg},
%             addressline={23 rue du Loess},
%             city={Strasbourg},
%             postcode={67034},
%             %state={},
%             country={France}}

\begin{abstract}
\subsection*{Hypothesis}
Although the dynamics of colloids in the vicinity of a solid interface has been widely characterized in the past, experimental studies of Brownian diffusion close to an air-water interface are rare and limited to particle-interface gap distances larger than the particle size. At the still unexplored lower distances, the dynamics is expected to be extremely sensitive to boundary conditions at the air-water interface. There, \textit{ad hoc} experiments would provide a quantitative validation of predictions.
\subsection*{Experiments}
Using a specially designed dual wave interferometric setup, the 3D dynamics of $9$ \textmu{}m diameter particles at a few hundreds of nanometers from an air-water interface is here measured in thermal equilibrium. 
\subsection*{Findings}
Intriguingly, while the measured dynamics parallel to the interface approaches expected predictions for slip boundary conditions, the Brownian motion normal to the interface is very close to the predictions for no-slip boundary conditions.
These puzzling results are rationalized considering current models of incompressible interfacial flow and deepened developing an \textit{ad hoc} model which considers the contribution of tiny concentrations of surface active particles at the interface.
We argue that such condition governs the particle dynamics in a large spectrum of systems ranging from biofilm formation to flotation process. 
\end{abstract}

%%Graphical abstract
%\begin{graphicalabstract}
%\includegraphics[width=13cm]{graphicalabstract.png}
%\end{graphicalabstract}

\begin{keyword}
%% keywords here, in the form: keyword \sep keyword
Brownian diffusion, Air-Water interface, Boundary conditions, Microscopy, Particle tracking.
%% PACS codes here, in the form: \PACS code \sep code

\end{keyword}

\end{frontmatter}

%% \linenumbers

%% main text
\section{Introduction}
The presence of an interface in an otherwise unbounded fluid breaks the isotropy and homogeneity of the space. For a particle moving in close vicinity of the interface these broken symmetries  manifest themselves in a tensorial and space dependent particle drag the values of which are determined by boundary conditions (BC) at the interface. 
Theoretical expressions for the translational mobilities are well known for sphere motion parallel and perpendicular to a planar interface in both cases of full slip and no-slip BC \cite{brenner1961slow,goldman1967slow,nguyen2004exact,villa2020motion} (see \ref{appendix_stateofart} for a report of the models). 
In the case of a water-air interface, the particle drag is expected to be governed by the free slip BC as predicted for free fluid surfaces \cite{brenner1961slow,nguyen2004exact}. 

For particle-interface gap distances $d$ larger or comparable to the particle radius $a$, the translational diffusive dynamics of micrometric particles along the directions parallel and perpendicular to a liquid-gas interface has been found in agreement with theoretical predictions taking into account full slip BC \cite{villa2020motion,wang2009hydrodynamic,benavides2016brownian,watarai2014experimental,boatwright2014probing}. 
Although these works seem to validate the use of the classical models of low Reynolds-number hydrodynamics, they did not probe the short gap distance regime ($d/a \ll 1$), where the dependence on the BC is expected to be stronger \cite{villa2020motion}.
Moreover, and more importantly, recent results by Maali \textit{et al.} \cite{maali2017viscoelastic} on the  perpendicular drag experienced by a bead forced to oscillate very close to an air-water interface contradict the classical model for a free fluid surface.
These authors observed indeed a frequency dependent drag, which at low frequency assumes values predicted for no-slip BC. 
The observed frequency dependence has been rationalized by Maali \textit{ et al.} considering the effect of surface active species at the interface, whose presence in real systems cannot be completely avoided because of the large air-water surface tension. These results show that a consensus in the field is still lacking.

A number of works have recently addressed the effect of surface active species on the dynamics at the interface \cite{arangalage2018dual,chisholm2021driven,molaei2021interfacial}, mainly focusing on how they affect the interfacial flow. The role of surface active agents on the Brownian motion of a particle very close to the interface, however, has never been considered and experimentally verified.
Far from being a mere fundamental physics topic, a full understanding of this process is demanded to properly describe and predict various natural and industrial processes. Prime examples include biofilm formation \cite{wotton2005surface}, bacterial dynamics \cite{bianchi20193d}, waste water treatment \cite{xing2017recent} and the formation of Pickering emulsions \cite{chevalier2013emulsions,alison20193d}.

In this Paper, we report on experimental measurements of the hydrodynamic mobility of spherical microparticles in the still unexplored close vicinity of a fluid interface. Analyzing the Brownian motion of the microparticles we are able to extract both parallel and perpendicular particle translational drag near the interface for gap distances in the range of $5\cdot10^{-3}a<d<1.5\cdot10^{-1}a$. 
For the same particles the measured dynamics along the two directions are compatible with predictions for different BC: slip for motion parallel to the interface and no-slip for the orthogonal one.

\section{Materials and Methods}

The experimental set-up and the tracking method are detailed in Ref. \cite{villa2020multistable}. The sample cell, sketched in Figure \ref{fig:sketch}b, is made by an hollow glass cylinder ($8$ mm inner diameter and $4$ mm height) closed on the top by a glass slide and open on the bottom. The cell is completely filled with water solutions made of deionized water (Millipore Milli-Q filtration system) and NaCl at different molar concentrations. The formed water-air interface is stabilized by the pinning of the contact line on the lower edge of the cylinder.

Polystyrene latex beads with nominal radius of $a=4.35 \pm 0.45$ \textmu{}m and refractive index $n_p = 1.59$ are added to the  solution at a concentration  $10^{-2}$ g L\textsuperscript{-1}. Because of gravity, particles sediment towards the air-water interface until they reach an equilibrium gap $d_0$ determined by the balance between gravity and DLVO forces.  The latter can be finely tuned by changing the amount of NaCl thus adjusting $d_0$ from $500$ nm (molar concentration of sodium chloride $\rho = 5 \cdot 10^{-6}$ M) to $50$ nm ($\rho= 10^{-3}$ M). Such a system allows the study of particle dynamics in the particular interesting range of extremely low values of $d/a$, between $1.5 \cdot 10^{-1}$ and $5 \cdot 10^{-3}$, where the translational diffusion coefficient is expected to strongly depend on the distance \cite{brenner1961slow,goldman1967slow,nguyen2004exact}.
The concentration of particles at the interface is low enough to guarantee inter-particles mean distance of hundreds of micrometers. The hydrodynamic coupling between particles can then be safely neglected. 

The dynamics of microparticles near the interface is affected by drifts parallel to the air-water interface mainly due to air displacement close to the water surface. In order to reduce drift, the region between the sample
stage and the objective is enclosed in a chamber made of cellulose acetate and nitrile butadiene rubber, lined inside by an alluminium foil to screen electrostatics charges. The chamber reduces evaporation and lowers drift from values of the order of $10$ \textmu{}m s\textsuperscript{-1} to $1$ \textmu{}m s\textsuperscript{-1}.

\subsection{Dual Wave Reflection Interference Microscopy}
\begin{figure}
\centering
\includegraphics[width=8.6cm]{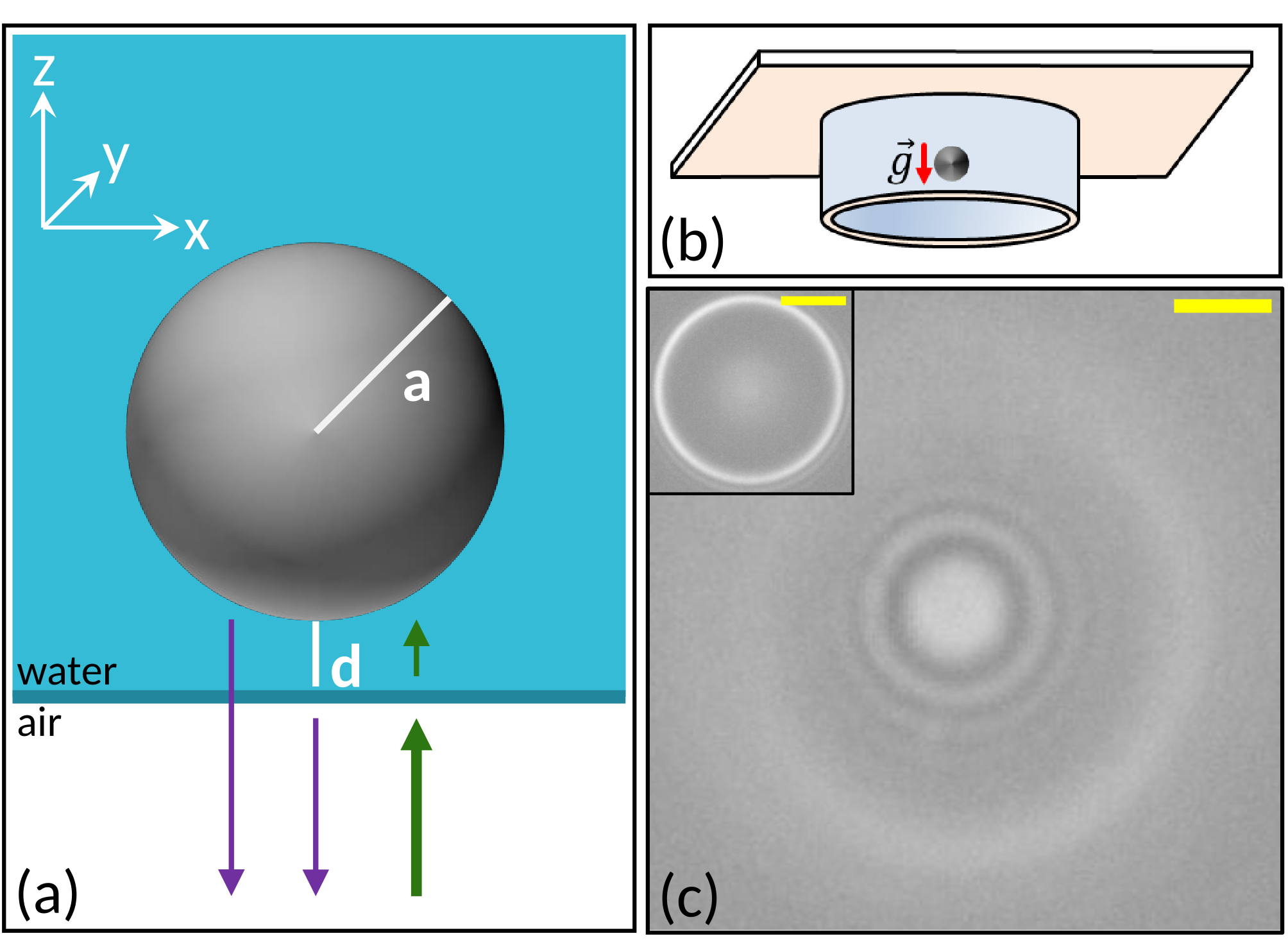}
\caption{Experimental setup. (a) Geometry of the interferometric pattern formation, the particle radius $a$ and the gap distance $d$ from the air-water interface. Green and purple arrows depict respectively the optical paths of the incoming and the reflected beams. (b) Sketch of the sample cell, filled with the water solution containing particles which sediment towards the interface because of gravity (red arrow). (c) Typical fringe pattern, obtained from the red channel of the camera focusing the microscope on the gap distance $d$ (yellow scale bar $1$ \textmu{}m). In the inset, image of the middle plane of the same bead from which particle radius can be measured (yellow scale bar $2$ \textmu{}m).}
\label{fig:sketch}
\end{figure}

In order to track the three-dimensional motion of particles dispersed in water and close to the interface with air a specifically designed Dual Wave Reflection Interference Microscopy (DW-RIM) setup has been employed \cite{villa2020multistable}. It consists of an interferometric system mounted on a custom inverted reflection microscope. The optical signal is generated in imaging configuration by the superposition of light originated from two narrow band LEDs, respectively centered at wavelengths $\lambda_{r}=625$ nm and $\lambda_{b}=505$ nm \footnote{\textit{Thorlabs collimated LEDs} M625L3-C2 and M505L3-C2.}. Each beam is reflected both from the particle and from the interface (see Figure \ref{fig:sketch}a) giving rise to an interference pattern made of concentric rings (see Figure \ref{fig:sketch}c for a typical example). Optimal imaging of the pattern is gathered when the microscope focal plane lies within the gap distance. The center and phase of the pattern respectively give the position of the particle in the plane of the interface $\left( x_c,y_c \right)$ and the particle-interface gap distance $d$, \textit{i.e.} the particle surface-to-interface minimal distance (as depicted in Figure \ref{fig:sketch}a). 

The choice to use two different sources, simultaneously recorded in distinct channels of an RGB camera \footnote{\textit{Basler dart camera} daA1280-54uc.}, is made to unambiguously determine the particle-interface gap distance \cite{villa2020multistable}.
From each wavelength, indeed, particle-interface gap distance can be measured up to an additive factor as $d_k=d_{k,0}+\tilde{m}_k\lambda_k/\left(2n\right)$, with $\tilde{m}_k$ being an integer number and $k=r$ and $b$ for wavelengths $\lambda_{r}$ and $\lambda_{b}$ respectively. 
The values of $\tilde{m}_r$ and $\tilde{m}_b$ are determined as the ones for which $d_r=d_b=d$. Residual ambiguity on $d$ due to matching periodicity of $d_r$ and $d_b$ as $\tilde{m}_r$ and $\tilde{m}_b$ increase is prevented by the limited coherence length of the employed light sources (in water equal to $13$ \textmu{}m and $5$ \textmu{}m for red and blue LEDs respectively).
Experimental interference patterns for beads at a distance of more than $1$ \textmu{}m exhibit indeed a sensibly lower contrast than the one reported in Figure \ref{fig:sketch}c, and they almost disappear for distances of more than $4$ \textmu{}m. All particles measured at the equilibrium gap distance show a well-contrasted interference pattern, thus limiting for them the possible values of $\tilde{m}_k$ in the range $0-5$. With such a limited set of values, cross-check between the two interference patterns allows an unambiguous determination of $d$.
A validation of the described method comes from the observation of trajectories of beads followed from sedimentation to $d_0$ and than observed while they are adsorbed at the interface. % [REF NEW FIGURE].
The latter event occasionally occurs in experiments at the higher explored salt concentrations of the water solution. Particle interface breaching can be unambiguously determined as the contact line is clearly visible on the image. On these datasets $d\left(t\right)$ is therefore directly measured from the single interference pattern. These trajectories can be therefore used to validate the procedure for particle-interface gap distance measurement from the cross-check of the two interference patterns.

For the analysis, a custom tracking algorithm has been specifically developed to determine both $\left( x_c,y_c \right)$ and $d$ with respective resolutions of $50$ nm and $10$ nm. 
Using this method we measured the trajectories at the equilibrium of $65$ different polystyrene particles at $35$ frames per second collecting and analyzing $5000$-$10000$ frames per particle.

The radius of each particle has been measured moving the focal plane to the particle middle-plane, where a bright circular corona is visible whose diameter corresponds to the particle size (inset of Figure \ref{fig:sketch}c). Corona radius is recovered from a parabolic fit of its azimuthal average with a precision of the order of $50$ nm, due to the error in localizing the center of the corona. The precise relation between the corona diameter and real particle size has been statistically investigated. The distribution of all the corona sizes measured during the experimental campaign has been compared with the distribution of particles radii obtained from bright field images of dried sample. To obtain the latter, a drop of solution containing particles has been deposited on a glass slide and dried in an oven. Clusters of touching particles formed and particles radii have been obtained measuring the particles center-to-center distance. Comparison between the distributions revealed that measured corona diameter is systematically larger than particle real size by a factor $1.04$. We checked the absence of a significant dependence of this factor upon particle radius repeating the characterization on a different bunch of microparticles with a nominal distribution diameter of $6.2\pm0.4$ \textmu{}m. We recovered for them the same corrective factor $1.04$.
For the data reported in the text we therefore obtain the single particle radius by dividing by this factor the radius of the corona imaged in reflection.

\begin{figure}
\centering
\includegraphics[width=8.6cm]{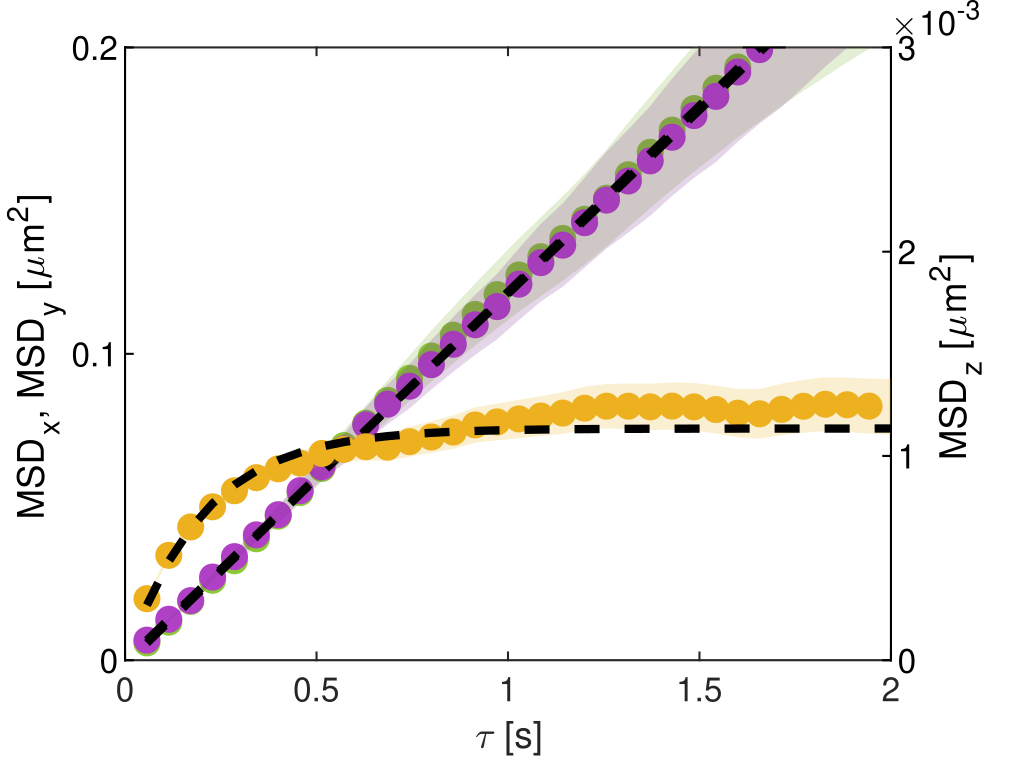}
\caption{Typical mean square displacements of a particle $a=4.48\pm0.06$ \textmu{}m at an average distance $d=260\pm10$ nm from the interface. The ordinate axis on the left corresponds to $\text{MSD}_x$ (green) and $\text{MSD}_y$ (purple), while the one on the right to $\text{MSD}_z$ (yellow). Shadows delimit experimental incertitude while black dashed lines represent data best fits. From the fits of the parallel dynamics compatible diffusion coefficients $D_x=\left(6.07\pm0.05\right)\cdot 10^{-14}$ m\textsuperscript{2}s\textsuperscript{-1} and $D_y=\left(6.00\pm0.04\right)\cdot 10^{-14}$ m\textsuperscript{2}s\textsuperscript{-1} are found. From $\text{MSD}_z$ fit, parameters $\omega_0 = \left(4.12\pm0.03\right)\cdot 10^3$ s\textsuperscript{-1} and $\omega_\mu = \left(2.3\pm0.2\right)\cdot 10^6$ s\textsuperscript{-1} are obtained, which corespond to a $0.27\pm 0.02$ s characteristic time for reaching the plateau and to a diffusion coefficient $D_z=\left(0.23\pm0.02\right)\cdot 10^{-14}$ m\textsuperscript{2}s\textsuperscript{-1}.}
\label{fig:msd}
\end{figure}
 
The equilibrium position the particles reach after sedimentation has a transient nature. Because of thermal fluctuations, indeed, beads can eventually get absorbed at the interface \cite{kaz2012physical} or move between two metastable equilibrium positions \cite{villa2020multistable}. These dramatic events have a clear signature in the trajectory orthogonal to the interface \cite{villa2020multistable}. In order to avoid artifacts in the analysis, we therefore visually inspected the orthogonal dynamics of all the recorded particles and excluded from the analysis the beads moving between metastable equilibria or adsorbed at the interface during the experimental time window\footnote{For more details on the nature of the metastable equilibrium, we refer to the work in Ref. \cite{villa2020multistable}, where data obtained with the same setup of the present work are used to investigate the DLVO interaction between the particles and the interface. Briefly, existence of metastable solutions is due to the eventual presence of nanometric air-bubbles stuck on the particle surface that, depending on the bead orientation with respect to the interface, could affect the DLVO interaction between the particle and the interface. 
Beads considered in the present work all lie in the first equilibrium position, where air bubbles are far from the interface and thus do not affect the dynamics. It is important to note that only a small fraction of the considered tracked particles present surface air-bubbles and that their dynamics around the first equilibrium position is indistinguishable from the one of the beads without air-bubbles.}.% [ADD FIGURE WITH d(t) FOR BEADS AT EQU, AT TRANSITION BETWEEN METASTABLE EQ, AND CROSSING INTERFACE ]}
 
 \subsection{Mean Square Displacement Analysis}

 For each translational degree of freedom the Mean Square Displacement at lag time $\tau$
 \begin{equation}
     \text{MSD}_k\left(\tau\right)=\left<\left[k\left(t+\tau{}\right)-k\left(t\right)\right]^2\right>_t,
 \end{equation}
 where $k$ is either $x=x_c$, $y=y_c$ or $z=d$, is obtained after removing from $x$ and $y$ the ballistic contribution due to drift.
On experimental time scales, particles drift parallel to the interface at constant velocity. Drift correction is therefore made before computing the MSDs by subtracting from $x\left(t\right)$ and $y\left(t\right)$ quantities $v_xt$ and $v_yt$ respectively, where $v_x$ and $v_y$ are obtained as the coefficient of the linear fit of $x$ and $y$ over time.
 Typical experimental $MSD$s for the Brownian motion of particles fluctuating around the equilibrium gap distance $d_0$ are reported in Figure \ref{fig:msd}. The linearity and the equal slope of $\text{MSD}_x$ and $\text{MSD}_y$ are respectively signatures of free diffusion and motion isotropy in the plane parallel to the interface. 
 Accordingly, the diffusion coefficients $D_x$ and $D_y$ are obtained by linearly fitting $\text{MSD}_x$ and $\text{MSD}_y$ as $\text{MSD}_k=2D_k\tau$. 
 The $\text{MSD}_z$ shows a different behavior, with an increasing trend at short lag-times followed by a plateau at larger lag-times, pointing out the confinement in $z$ due to the potential well rising from gravity and the interaction with the interface. 
 At short lag-times, the dynamics of the particle is close to the one of a free particle. As the particle is confined in an energy well around the potential minimum, however, the explored distances are bounded due to the increasing restoring forces which gives rise to a plateau at larger lag-times.
  In order to fit the $\text{MSD}_z$ we use the analytical solution of the Langevin equation in the overdamped regime assuming the presence of a harmonic potential \cite{wang1945theory,li2010measurement}:
  
 \begin{equation}
     \text{MSD}_z\left(\tau\right) = \frac{2k_BT}{m{\omega_0}^2}\left[1-e^{-\omega_{\mu}\tau}\left(\cosh{\tilde{\omega}\tau} + \frac{\omega_{\mu}}{\tilde{\omega}}\sinh{\tilde{\omega}\tau}\right)\right],
%\label{eq_msd_harmonic_potential}
\end{equation}
where $k_BT$ is the product of temperature and Boltzman constant, $m$ is the particle mass, % corrected by buoyancy,
$\omega_0$ is the characteristic frequency of the potential well, $\omega_\mu=k_BT/2D_zm$ and $\tilde{\omega}=\sqrt{\omega_\mu^2 - \omega_0^2}$. 
 Low experimental values of $\omega_0$ compared to $\omega_\mu$ safely allow the overdamped approximation $\tilde{\omega}\simeq \omega_\mu - \omega_0^2/2\omega_\mu$, thus resulting in the $\text{MSD}_z$ expression:
 
  \begin{equation}
     \text{MSD}_z\left(\tau\right) = \frac{2k_BT}{m{\omega_0}^2}\left(1-e^{-\frac{\omega_0^2\tau}{2\omega_\mu}}\right).
\label{eq_msd_harmonic_potential}
\end{equation}
 
  The validity of the harmonic approximation for the DLVO and gravitational potential has been verified through dedicated simulations comparing Brownian dynamics in harmonic and non-harmonic potential wells.
 Results show that, for the considered potentials and at experimental temperatures, non-harmonic contributions to the  diffusion coefficient can be neglected. %\cite{villa2021preparation}.

\section{Results and Discussion}
\subsection{Experimental Diffusion Coefficients}
\begin{figure}
\centering
\includegraphics[width=8.6cm]{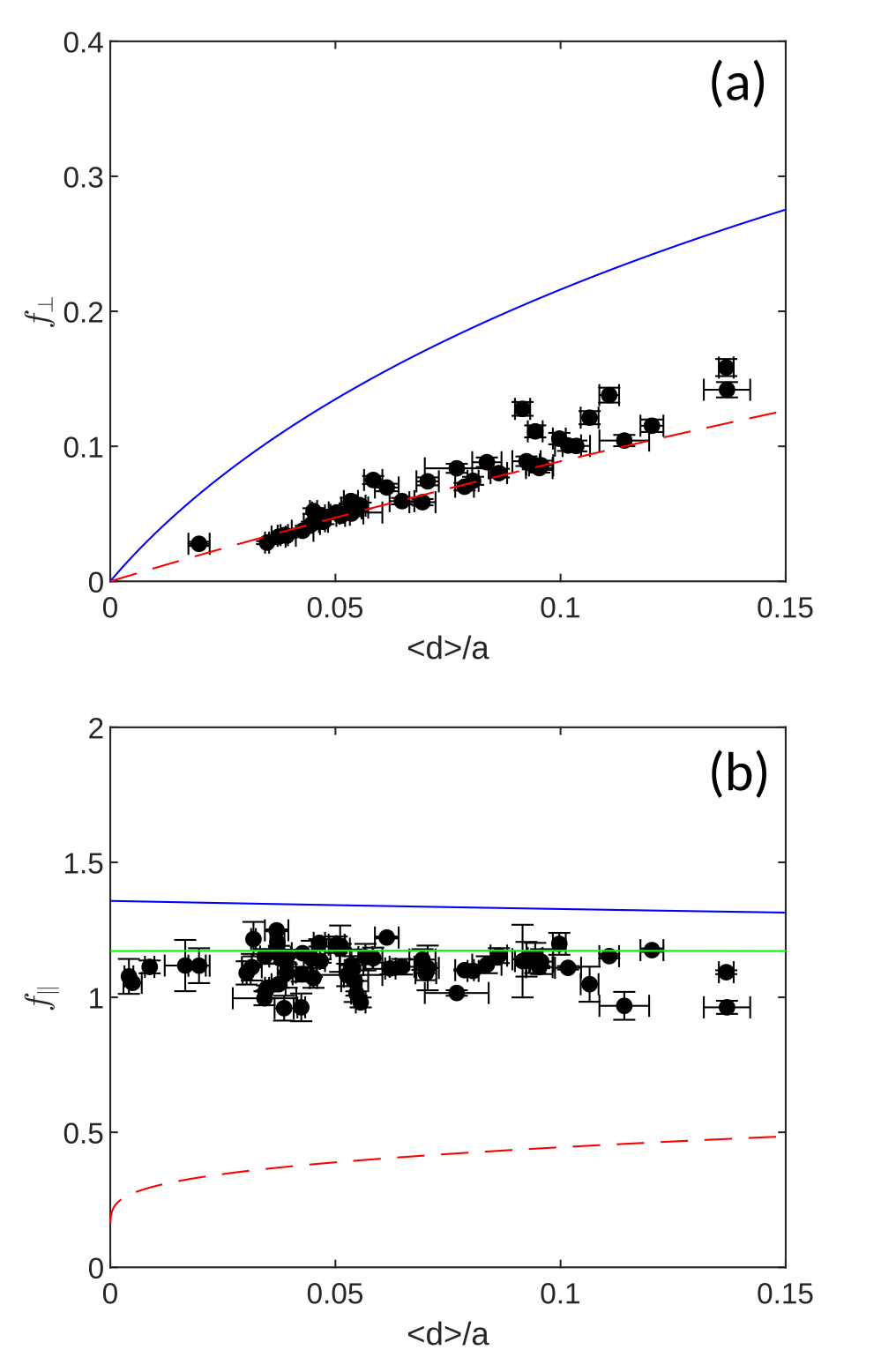}
\caption{Measured dimensionless diffusion coefficients (black points) perpendicular (a) and parallel (b) to the air-water interface as a function of $<d>/a$. Data refer to different particles and fluid molarities. Theoretical predictions are also reported for no-slip (red line), full-slip (blue line) and, in (b), surface incompressibility boundary conditions (green line).}
	\label{fig:drag_all_data}
\end{figure}

In the following we consider the diffusion coefficient parallel and perpendicular to the interface $D_{\parallel} = \left(D_x+D_y\right)/2$ and $D_{\perp}=D_z$ respectively. 
Figure \ref{fig:drag_all_data}a and \ref{fig:drag_all_data}b report the ratios of the measured diffusion coefficients $f_{\parallel}=D_{\parallel}/D_0$ and $f_{\perp}=D_{\perp}/D_0$ respectively normalized by the bulk diffusion coefficient $D_0=k_BT/6\pi\eta a$ as a function of the normalized particle average distance from the interface $\left< d \right> /a$.
Experimental agreement with the expected value of $D_0$ has been verified with an \textit{ad hoc} bright field experiment. There, the Brownian dynamics of few beads in the focal plane has been measured in a density matched mixture of water and deuterium, in order to prevent sedimentation. Agreement of the measured diffusion with theoretical expectations $D_0=(5.6\pm 0.6)\cdot{}10^{-14}$ m\textsuperscript{2}s\textsuperscript{-1} has been found within the experimental incertitude: $(7\pm 2) \cdot{}10^{-14}$ m\textsuperscript{2}s\textsuperscript{-1}.

In Figures \ref{fig:drag_all_data}a and \ref{fig:drag_all_data}b each point corresponds to a different particle. Measured particles come from different bunches and have been measured in different days. Measurements performed on the same particle for more than $200$ s and on different particles at the beginning and at the end of a single experiment (time scale of few hours) do not reveal any aging of the interface.
In the plots the theoretical predictions for $f_{\parallel}$ and $f_{\perp}$ for both slip and no-slip BC (\ref{appendix_stateofart}) are also reported. 
Experiments and theory show a dependence of $f_{\perp}$ and, to a lower extent, $f_{\parallel}$ on the particle-interface distance. 
Note that such a dependence in principle affects the above $\text{MSD}$s analysis that has been performed assuming a constant drag coefficient all along the trajectory.
The relevance of this effect has been addressed by taking the weighted average of the theoretical predictions $\left<P\left(d\right)f_i\left(d\right)\right>_d$, where $i$ indicates either motion parallel and normal to the interface and $P(d)$ is the  probability distribution function of $d$ around the potential minimum. We found negligible deviations between the calculated $\left<P\left(d\right)f_i\left(d\right)\right>_d$ and the measured $f_i\left(\left<d\right>\right)$, thus validating our analysis.

 From Figure \ref{fig:drag_all_data}b it can be seen that the experimental $f_{\parallel}$ approaches the expected predictions for full slip BC (blue line), although remaining systematically lower by on average $14\%$. On the other side,  $f_{\perp}$ deviates from expectations and follows instead the predictions for no-slip BC (red dashed line in Figure \ref{fig:drag_all_data}a).

\subsection{Possible Sources of Additional Dissipation}
Theoretical predictions for slip and no-slip BC are made considering spheres with perfectly flat surface. AFM measurements of the used particles revealed an average roughness of the order of $5$ nm with occasionally few isolated imperfections of tens of nanometers \cite{villa2020multistable}. 
Importantly, no differences in the dynamics has been observed between beads with and without large imperfections\footnote{Except for few cases where rotational lock occurs because of DLVO interaction (for details, see reference \cite{villa2020multistable}). These particles have been excluded from the present analysis.}. Moreover, an alteration of the dynamics due to deviation from ideal surface flatness is expected to be more relevant at shorter distances (comparable with the surface roughness) and negligible at larger distances because of the aforementioned agreement of the bulk drag with theoretical expectations. 

A possible further source of dissipation are electrokinetic effects, rising from the particle and air-water interface double layer perturbation induced by their relative motion. Electrokinetic contribution is however expected to be negligible in the present system. 
On the basis of existing theories, indeed, electroviscous drag in the bulk is expected to be for the present system $4$ order of magnitude lower than hydrodynamic one \cite{ohshima1984}. As a particle approaches the interface, electroviscous drag increases approximately as $\left(d/a\right)^2$ \cite{tabatabaei2006,tabatabaei2010}, against the trend $\sim d/a$ of the hydrodynamic one. The stronger spatial dependence of the electroviscous drag, however, is not sufficient to make its contribution significant in the range of $d/a$ values explored in the present work, as it remains $10^2$ times lower than viscous drag (see \ref{appendix_electrokinetics}).

\subsection{Surface Active Species Contribution}
In order to rationalize the experimental results one needs to consider that due to the particle proximity to the interface, the particle dynamics is strongly influenced by the hydrodynamic boundary conditions and especially by the nature of the 2D  flow at the interface. 
The experimental results suggest that the 2D hydrodynamic flow at the interface induced by the particle movement does not follow the one predicted by hydrodynamics theory of a pure fluid interface. In particular, the data agreement with the no-slip BC model for particle movement perpendicular to the interface hints to the absence of interfacial flow, whereas the slip conditions for the parallel dynamics is the signature of its presence. These observations suggest some peculiar boundary conditions for the interfacial flow that neither a pure fluid interface nor a solid boundary can fulfill. Since the flow field at the interface $\vec{v}_s$ generated by a particle moving perpendicularly to the interface is purely radial, a sufficient condition to prevent surface flow in this case would be  $\nabla\cdot\vec{v}_s = 0$. 
Note that for a generic motion of the particle with respect to the interface such condition still allows an interfacial flow (\textit{e.g.}, a dipolar flow field).

 A similar boundary condition has been recently invoked by Maali \textit{et al.} to explain the measured drag of a bead glued on an AFM tip and forced to oscillate orthogonal to an air water interface \cite{maali2017viscoelastic}. In such experiment, the decrease of the forcing frequency $\omega$ corresponds to an increased measured drag which goes from the one corresponding to full slip to the one equivalent to no-slip BC. This frequency dependence reveals the existence of an interface characteristic frequency $\omega_c$ which is strictly linked to the presence of surface active species at the interface: $\omega_c = \frac{c_0 k_B T}{8 \eta a}$, where $c_0$ is the surface concentration of active species and $\eta$ the water viscosity. 
Even if stringent precautions are taken, low concentrations of surface active species (and molecular ionic species as well) are indeed always present and need to be taken into account when dealing with the dynamics close (or at) an air-water interface \cite{maali2017viscoelastic,uematsu2019impurity,manor2008hydrodynamic,koleski2020azimuthal}.
Accordingly to the model in Ref.\cite{maali2017viscoelastic}, if the forced bead oscillation frequency is low enough compared to $\omega_c$, bead dynamics timescale is lower than the one of surface active agents, which thus instantaneously react to the bead-induced surface flow through Marangoni stress, preventing concentration gradient formation. This results in the condition of surface incompressibility $\nabla\cdot\vec{v}_s = 0$ equivalent, for dynamics normal to the interface, to no-slip boundary conditions. In the opposite limit, when $\omega \gg \omega_c$, the surface active species are not able to follow the rapid movement of the bead: they only 'feel' its average which is zero. Consequently no stresses build up and full slip BC apply.

 By fitting experimental data collected in different days, Maali \textit{et al.} obtained an average surface active species concentration $c_0=75\cdot 10^{15}$ m\textsuperscript{-2} with a $16\%$ daily variation. It is important to notice how such a concentration, corresponding to an average area per molecule of $13$ nm\textsuperscript{2}, is far too low for being detectable through surface tension measurements \cite{kaganer1999structure}. Moreover, the low variability of concentration points out to a relatively robust characteristic frequency $\omega_c$ where the transition from slip to no-slip BC is expected.
 
If the model proposed by Maali \textit{et al.} is applied to our case, assuming a surface concentration of active species comparable to the one of Maali \textit{et al.} 
one finds $\omega_c \sim 10^{3}$ s\textsuperscript{-1}. It is worth noting that our low sampling frequency of $\nu=17.5$ s\textsuperscript{-1} $\ll \omega_c$ (corresponding to $35$ fps) filters out all the high-frequency components of the Brownian motion 'feeling' slip BC. As a consequence we expect that the measured behavior corresponds to an incompressible flow at the interface and no-slip BC. 

This argument applies also to the motion parallel to the interface. Numerical simulations by B{\l}awzdziewicz \textit{et al.} \cite{blawzdziewicz2010motion} of a colloidal particle translating parallel to an incompressible air-water interface, indeed, predict a diffusivity reduction of the order of $12\%$ with respect to slip BC for the range of relative gap distances explored in our work.
As it can be seen in Figure \ref{fig:drag_all_data}b, a satisfactory agreement is found between experimental values of drag parallel to the interface and B{\l}awzdziewicz \textit{et al.} simulations (green line). 
This represents the first experimental evidence on the effect of surface incompressibility for a particle motion close and parallel to a fluid interface.

The collapse on the same curve, within the experimental variability, of all the measured particles, whose trajectories have been measured in different days and at different times from the sample preparation, suggests that the concentration at the interface of surface active agents is rather constant and that the related adsorption process occurs within the first few minutes after the sample preparation.

\begin{figure}
\centering
\includegraphics[width=8.6cm]{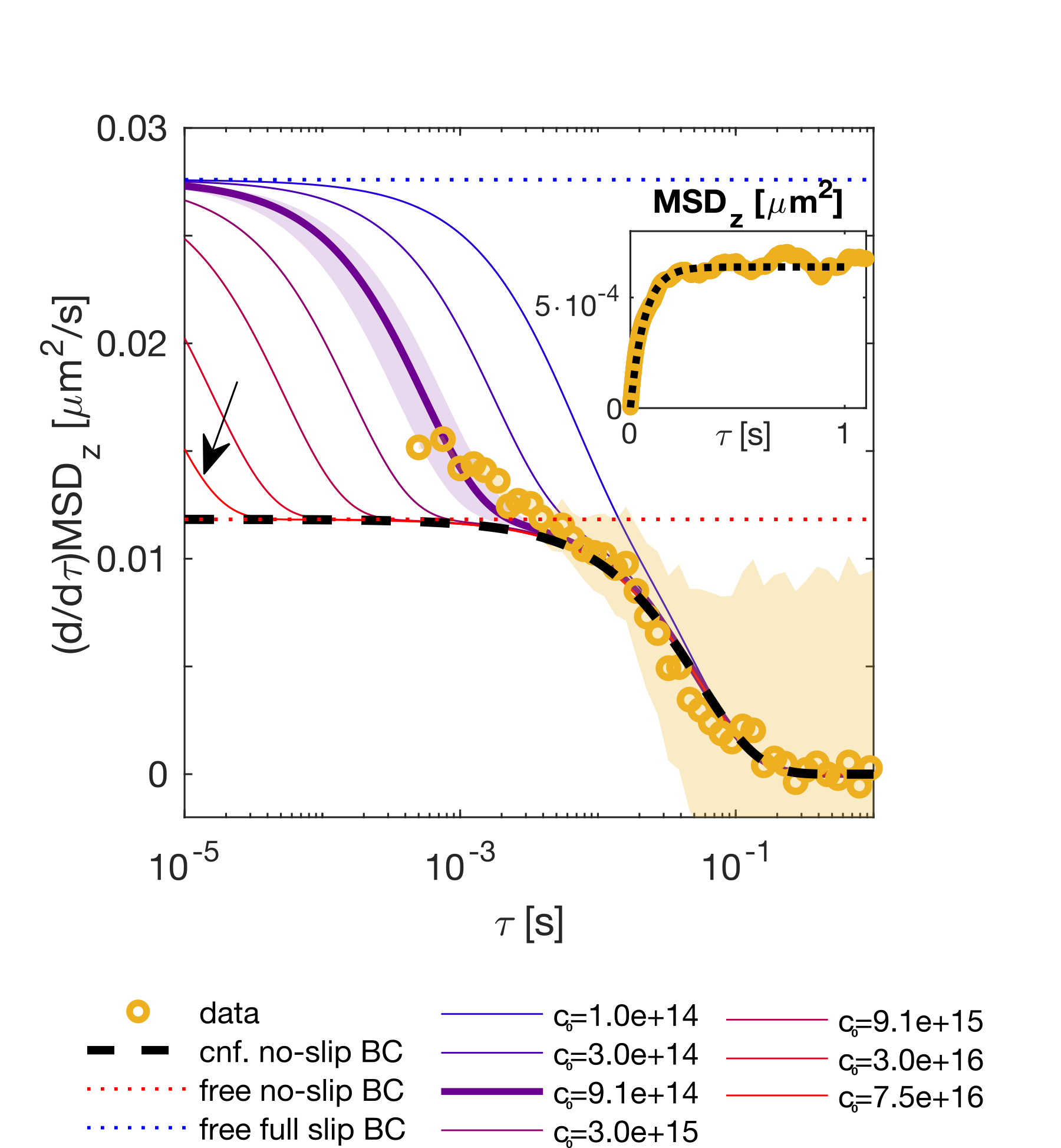}
\caption{Time derivative of $\text{MSD}_z$ for a bead recorded at $4000$ frames per second (yellow points). Particle has measured radius $a=4.42$ \textmu{}m and equilibrium gap distance $d_0=550$ nm. The sodium chloride molar concentration is $\rho=5\cdot{}10^{-6}$ M.
Shadowed yellow region includes experimental error intervals. The corresponding $\text{MSD}_z$ is reported in the inset (yellow points) with the no slip BC prediction (dashed black line). In the main plot, dotted lines represent the expected values for a free Brownian motion orthogonal to an interface with full-slip (dotted blue line) and a no-slip (dotted red line) BC, while black dashed line is the derivative of the $\text{MSD}_z$ prediction with no slip BC. Continuous lines represent predictions for a Brownian confined motion with a surface concentration $c_0$ of surface active species, with $c_0$ ranging from $10^{14}$ m\textsuperscript{-2} to  $75\cdot 10^{15}$ m\textsuperscript{-2} with a logarithmic spacing. The ticker continuous line corresponds to the best fitting $c_0=9\cdot 10^{14}$ m\textsuperscript{-2} and the shadowed purple region highlights its incertitude range of concentrations. The arrow points to the prediction for $c_0=75\cdot 10^{15}$ m\textsuperscript{-2}.}

\label{fig:phantom}
\end{figure} 

\subsection{Frequency Dependent Mobility Normal to the Interface}
In order to have further confirmations and verify whether the high frequency limit of the model of Maali \textit{et al.} can be reached in a thermally driven dynamics, we have explored the fast dynamics of the beads by recording particle trajectories at $4000$ fps\footnote{\textit{Phantom} VEO $1310$ camera.}. 
In Figure \ref{fig:phantom} it is reported an example of the first derivative of the $\text{MSD}_z$ with respect to the lag time as a function of $\tau$. Also shown in Figure \ref{fig:phantom} are the predicted values of the derivative for a purely diffusive motion for no-slip (red dotted line) and slip (blue dotted line) BC. The measured $\text{MSD}_z$ slope is not constant with $\tau$ but is lower than theoretical predictions at large $\tau$, it then crosses the expected value of no slip BC and tends to intermediate values between slip and no-slip BC at very low $\tau$. The behavior at large $\tau$ is known to be dominated by confinement effects due gravity and to the interaction potential between the particle and the interface. As a guideline, it is also reported as a black dashed line in Figure \ref{fig:phantom} the $\tau$ derivative of equation \ref{eq_msd_harmonic_potential} for a bead diffusing in a harmonic potential of stiffness $K_\omega=m\omega_0^2$. There, the diffusion coefficient $D_\perp$ is evaluated from no-slip BC predictions and is thus fixed once the bead radius and the particle-interface distance are known\footnote{Since the fast camera has a monochromatic sensor, these particular measurements are performed only with the red source beam. The average gap distance $\left<d\right>$ is therefore known up to an additive factor as $\left<d\right>=\left<d_r\right>+\tilde{m}\lambda_r/2n$ , where $\tilde{m}=0,1,...$ and $d_r$ is the distance directly recovered from the red channel interference pattern. An additional step is thus required to obtain the distance. Determination of $\tilde{m}$ is made by first fitting the experimental points reported in Figure \ref{fig:phantom} with the derivative of eq. \ref{eq_msd_harmonic_potential} keeping $d$ as the only free fitting parameter ($\omega_0$ is measured as described in the text). Fit is made for $\tau>0.1$ s in order to exclude small timescales where the frequency dependence of the diffusion parameter is expected to be more relevant. The obtained gap distance value $d_{fit}$ is then compared with the possible values given by $\left<d_r\right>+\tilde{m}\lambda/2n$ and $\tilde{m}$ is thus univocally determined as the integer minimizing the difference between the discrete values and the fitted distance.}.
The potential stiffness is obtained from a parabolic fit of the potential  $\Delta{}U\left(d\right)= -k_BT \cdot{}ln\left[P\left(d\right)\right]$ recovered from data. As it can be seen in the inset of Figure \ref{fig:phantom}, there is a good agreement between theoretical (with no free parameters) and measured $\text{MSD}_z$. The agreement is also very good between their derivatives for $\tau>0.03$ s, while at short lag times theoretical prediction extrapolates to the value of the diffusion coefficient for no-slip BC while experimental points continue increasing as $\tau$ decreases.
The effect of bead confinement, while reducing the effective diffusivity, cannot in any way cause a diffusion larger than the one expected in a free Brownian motion\footnote{Please note that, although we are considering very small time scales, we are still far away from the value $\tau_p$ at which the thermal dynamics is expected to became ballistic: $\tau_p = m\left(6\pi \eta a\right)^{-1}<3\cdot10^{-6}$ s, where $m$ is the particle mass.}. 
Data at high frequencies (short lag time) larger than the asymptotic limit means therefore that the effective diffusion coefficient is not constant but increases, in qualitative agreement with the model of Maali \textit{et al.} 

In order to be more quantitative we derived an analytical expression for the MSD considering the effect of the presence of a surface concentration $c_0$ of surface active species on the dynamics of a sphere orthogonal to the interface. Details of the model are reported in \ref{appendix_model}.
Briefly, building up on Maali \textit{et al.} model, we recover the frequency dependent drag coefficient for oscillatory dynamics. We then generalized the expression beyond the lubrication approximation including Brenner's general full slip and no-slip expressions (Eq. \ref{form:_f_perp_LS}) as the limit for $\omega\gg\omega_c$ and $\omega\ll\omega_c$ respectively.
Using a generalized Langevin equation also including the potential well, we finally derive an expression for the MSD in the overdamped limit including the frequency dependent drag:

\begin{equation}
	\mathrm{MSD} = Z
	\left( 2 \Lambda + 4 \Gamma - X_+ e^{-\omega_+ \tau} - X_- e^{-\omega_- \tau} \right) ,
	 \label{form:MSDcomplete}
\end{equation}
where parameters $Z$, $\Gamma$, $\Lambda$ and $X_\pm$ and frequencies $\omega_\pm$ are defined in \ref{appendix_model}. Frequencies $\omega_-$ and $\omega_+$ are related to the characteristic times of the potential well and of the surface active species reaction to the surface flow respectively.
Please note that the MSD in Eq.~\ref{form:MSDcomplete} includes contribution of both viscous and elastic response of the interface. For the present physical system and in the explored range of frequencies, however, the elastic response is much smaller than the viscous one \cite{maali2017viscoelastic} and a proper characterization of the two separate contributions can hardly be made. 
Here the displacements resulting from the whole visco-elastic response of the interface are therefore studied, leaving to future studies a deeper investigation on the complex modulus building on the model in \ref{appendix_model}.

In Figure \ref{fig:phantom} are reported different plots of the time derivative of Eq. \ref{form:MSDcomplete}, \textit{i.e.}:

\begin{equation}
	\frac{\partial{}}{\partial{}\tau} \mathrm{MSD} = Z
	\left( \omega_+ X_+ e^{-\omega_+ \tau} + \omega_- X_- e^{-\omega_- \tau} \right) ,
	 \label{form:derMSDcomplete}
\end{equation}
for different values of $c_0$, ranging from $10^{14}$ m\textsuperscript{-2} to  $75\cdot 10^{15}$ m\textsuperscript{-2} with a logarithmic spacing\footnote{All other parameters are fixed to the measured experimental values.}. As it can be seen, for surface active species concentrations of the order of $10^{14}$ m\textsuperscript{-2}  the particle dynamics is slightly affected by the resulting Marangoni effect and the $\text{MSD}$ time derivative decays exponentially from the full slip diffusion prediction (blue dotted line) to zero because of the bead confinement in the potential well. As $c_0$ increases, the two exponential decays become distinguishable, with a first decay from the full slip to the no-slip nominal values of $2D_\perp$ followed at the characteristic potential well time scale by a second decay to zero. Continuing increasing $c_0$, the model prediction becomes closer and closer to the confined dynamics expected orthogonal to a no-slip interface. Consequently, if the lowest experimentally accessible time scale is larger than $1/\omega_+$, like for the particles of Figure \ref{fig:drag_all_data}, the measured $\text{MSD}$ is the same as the one for a no slip BC.

The ticker violet line at a concentration $c_0= \left(9 \pm 4\right) \cdot 10^{14}$ m\textsuperscript{-2} corresponds to the best fit of the experimental data while the leftmost model curve (pointed by the black arrow) corresponds to the average surface active species concentration found by Maali \textit{et al.} The latter is almost two orders of magnitude larger than the one obtained for the present system. The lower degree of purity in the Maali \textit{et al.} system can be explained by the fact that an AFM tip is immersed in the solution, thus making possible an additional source of pollution to the system.

\section{Conclusions}
In conclusion, using specifically designed Dual Wave Reflection Interference Microscopy (DW-RIM) we achieve to measure the still unexplored dynamics of the three-dimensional movement of thermal spherical particles in close vicinity of an air-water interface. In such conditions the particle dynamics is expected to be extremely sensitive to the boundary conditions at the interface. Intriguingly, measured dynamics parallel and normal to the interface are close to predictions for slip and no-slip BC respectively.
 Such puzzling results can be rationalized in the theoretical framework of Maali \textit{et al.} \cite{maali2017viscoelastic} and B{\l}awzdziewicz \textit{et al.} \cite{blawzdziewicz2010motion} that considered an incompressible flow at the interface. For a radial flow at the interface, induced by a particle orthogonal movement, the incompressibility condition is equivalent to a no-slip boundary \cite{maali2017viscoelastic}. 
 Conversely, for motion parallel to the interface this condition only yields  a slight deviation from the slip BC predictions \cite{blawzdziewicz2010motion}.
Both these models are in agreement with our experimental findings.

We also developed a theoretical model which takes into account the effect of surface active agents on Brownian dynamics close and orthogonal to the interface. The agreement between the model and data acquired at high frame rate is remarkably good. The low value of the obtained best fitting surface active species concentration justifies the ideal gas approximation. It also explains why at typical acquisition frame rates the surface active species effect on the sphere dynamics orthogonal to the interface cannot be distinguished from the no slip BC case. The agreement of the measured diffusion coefficients parallel to the interface with the surface incompressibility prediction by B{\l}awzdziewicz \textit{et al.} \cite{blawzdziewicz2010motion} also supports this interpretation.

Future works should clarify the nature of the surface active species. Surface incompressibility can originate from low concentration of ionic species (such as bicarbonate) at the interface, due to the equilibrium between water and atmosphere \cite{yan2018central}. An illuminating test in this direction would be a repetition of our experiment in a controlled-atmosphere environment. Additionally, controlled addition of either insoluble and soluble surfactants at the interface \cite{kaganer1999structure} would shed more light on the phenomena described in the present paper, also exploring eventual effects of surfactants desorption rate on the bead dynamics \cite{villa2020motion,mucic2011dynamics}. The extremely low concentrations of the present and Maali \textit{et al.} works as deduced from the models makes it difficult their direct measurement by the means of standard techniques. In this sense, further investigations in this direction could proceed by combining optical tweezers \cite{shlomovitz2014probing,boatwright2014probing} with bead tracking in the close vicinity of the interface, in regimes inaccessible with standard methods \cite{villa2020motion}. We believe that our technique implemented with an optical tweezer with intermittent and tracking-synchronized trap will open new avenues of research in interfacial rheology.

Recently, the effect of surface incompressibility upon interfacial dynamics at air-water interface is gaining rising interest \cite{manor2008hydrodynamic,chisholm2021driven,molaei2021interfacial}. Here we first show how it also affects the 3D dynamics of Brownian colloids in its vicinity. Practical declinations of the observed effect are expected to be important when describing biological phenomena and industrial processes involving micrometric objects approaching air-water interfaces. A bright example of this can be found in Bianchi \textit{et al.} \cite{bianchi20193d}, where surface active species presence at the air-water interface has to be taken into account to explain the dynamics of bacteria close to water surface.
In this framework future studies including non-spherical colloidal particles, also exploring surface incompressibility effect on rotational degree of freedom, would contribute to make an additional step forward in the direction of realistic natural systems.

%\begin{acknowledgments}
\section*{Acknowledgement}
The authors acknowledge financial support from the French Agence Nationale de la Recherche (Contract n° ANR-14-CE07-0039-SURFANICOL), and from the LabEx NUMEV (Contract n°AAP2014-2-044). We also acknowledge support from the Max Planck Center Twente for Complex Fluid Dynamics.
%\end{acknowledgments}
\section*{Author contributions}
\textbf{Stefano Villa:} Software, Methodology, Investigation, Formal analysis, Validation, Visualization, Writing (original draft and review editing).
\textbf{Christophe Blanc:} Methodology, Software, Resources, Writing (review editing).
\textbf{Abdallah Daddi-Moussa-Ider:} Methodology (theoretical modelling), Writing (original draft and review editing).
\textbf{Antonio Stocco:} Conceptualization, Resources, Writing (review editing), Supervision.
\textbf{Maurizio Nobili:} Conceptualization, Resources, Writing (original draft and review editing), Supervision, Project administration.

\appendix
\section{Exact Predictions for the Mobility of a Sphere Close to Full Slip and no-Slip BC Interfaces}
\label{appendix_stateofart}
Translational mobility for a sphere of radius $a$ orthogonal to an infinite plane interface was obtained by Brenner in 1961 \cite{brenner1961slow}. He determined the full series solution of the quasi-static Navier-Stokes equations for both full-slip and no-slip boundary conditions on the plane. By expressing the spatial dependency with the parameter $h' =  \cosh ^{-1} \left(1+d/a \right)$, where $d$ is the particle-interface gap distance, Brenner's exact solutions for the ratio $f_\perp$ between mobility $\mu_\perp$ orthogonal to an interface and the one in the bulk $\mu=1/\left(6\pi\eta{}a\right)$ are:

\begin{subequations}
\begin{align}
        \dfrac{1}{f_\perp^{FS}} = &\dfrac{4}{3} \sinh{h'} \sum_{n=1}^{\infty} \dfrac{n \left( n+1 \right)}{\left( 2n-1 \right)\left( 2n+3 \right)} \cdot&& \notag \\
        &\cdot \left[ \dfrac{4 \cosh^2\left(n+\frac{1}{2}\right)h' + \left(2n+1\right)^2\sinh^2 h'}{2\sinh{\left(2n+1\right)h'}-\left(2n+1\right)\sinh{2h'}}-1 \right],
        \end{align}
\label{form:_f_perp_LA}
\\
\begin{align}
\dfrac{1}{f_\perp^{NS}} = &\dfrac{4}{3} \sinh{h'} \sum_{n=1}^{\infty} \dfrac{n \left( n+1 \right)}{\left( 2n-1 \right)\left( 2n+3 \right)} \cdot&& \notag \\
&\cdot \left[ \dfrac{2\sinh{\left(2n+1\right)h'}+\left(2n+1\right)\sinh{2h'}}{4 \sinh^2\left(n+\frac{1}{2}\right)h' - \left(2n+1\right)^2\sinh^2 h}-1 \right],
\end{align}
\label{form:_f_perp_LS}
\end{subequations}

\noindent where $FS$ and $NS$ denotes, respectively, the full-slip and no slip BC at the interface cases. 
An analytical solution for the motion parallel to the interface is more difficoult to obtain due to the coupling between translational and rotational motion.
 The $d$ dependency of the drag on a finite size object, indeed, causes a non-zero torque on the particle upon the application of a parallel translational force. 
Goldman, Cox and Brenner (GCB) \cite{goldman1967slow} first found numerical solutions of Stokes equation for a sphere moving parallel to a solid wall. 
Nguyen and Evans \cite{nguyen2004exact} then found the corresponding exact numerical solutions for full slip BC on the plane. They also developed approximate solutions of both their and GCB's numerical solutions for the whole range of separating distances. 
Their approximated formula for the ratio  $f_\parallel$ between mobility $\mu_\parallel$ parallel to the interface and $\mu$ are:

\begin{subequations}
    \begin{align}
        \dfrac{1}{f_\parallel^{FS}} = \left \{ 1 + 0.498 \left \{ \ln{\left[ 1.207 \left(\dfrac{a}{d}\right)^{0.986} +1\right]} \right\}^{1.027}  \right \}^{0.979},
%\label{form:_f_parallel_LS}
\\
\dfrac{1}{f_\parallel^{NS}} = \left \{ 1 + 0.498 \left \{ \ln{\left[ 1.207 \left(\dfrac{a}{d}\right)^{0.986} +1\right]} \right\}^{1.027}  \right \}^{0.979}.
%\label{form:_f_parallel_LS}
    \end{align}
\end{subequations}

\section{Electrokinetic Effects}
\label{appendix_electrokinetics}

From the coupling of hydrodynamics and electric forces a number of phenomena arise which are grouped under the name of \textit{electrokinetic effects}. Since the present work addresses the hydrodynamic interaction between a sphere and an interface both negatively charged \cite{villa2020multistable}, these effects should be in principle taken into account.
Electrokinetics is indeed source of additional dissipations rising from the distortion of the double-layer due to the flow which, in turn, alters the local hydrodynamic flow around the particle leading to an increment of energy dissipation.

Effect of such electroviscous drag on a charged spherical particle of Zeta potential $\zeta_{p}$ translating in a bulk solution has been studied by Ohshima \textit{et al.} \cite{ohshima1984}. They consider the presence of two ionic species with valency $j_i$, diffusivity $D_{is,i}$ and bulk number density $n_i^0$, where $i=1,2$ refers to counter-ions and co-ions respectively. In the small Peclet number and Debye screening lengths limit they found for the electroviscous drag $\xi_{ev}$ the following expression \cite{tabatabaei2010}:

\begin{align}
\xi_{ev} = \dfrac{48\pi\left(\epsilon_0\epsilon\right)^2 \left(k_BT\right)^3}{e^4 \left(n_1^0j_1^2+n_2^0j_2^2\right)a} \left( \dfrac{G_p^2}{j_1^2D_{is,1}}+\dfrac{H_p^2}{j_2^2D_{is,2}} \right) \,,
\label{form:ev-drag_bulk}
\end{align}
where 

\begin{align}
G_{p} &= \ln{\left[\dfrac{1+\exp\left(\dfrac{j_1 e \zeta_{p}}{2k_BT}\right)}{2}\right]}
\label{form:ev-drag_bulk_defGH1}
\end{align}
and 

\begin{align}
H_{p} &= \ln{\left[\dfrac{1+\exp\left(\dfrac{j_2 e \zeta_{p}}{2k_BT}\right)}{2}\right]}.
\label{form:ev-drag_bulk_defGH2}
\end{align}

Since the water solutions considered in the present work are obtained dissolving Sodium Chloride in deionized water, we can restrict here to the case of symmetric electrolytes ($n_1^0=n_2^0=:n^0$ and $D_{is,1}=D_{is,2}=:D_{is}$) of valency $j=1$. Accordingly, $\xi_{ev}$ can be rewritten as

\begin{align}
\xi_{ev} = 24 C \left( G_p^2+H_p^2\right) \,,
\label{form:ev-drag_bulk2}
\end{align}

\noindent where $C:= \pi\left(\epsilon_0\epsilon\right)^2 \left(k_BT\right)^3/e^4n^0aD_{is}$.

In the same limits of Ohshima \textit{et al.}, Tabatabaei \textit{et al.} \cite{tabatabaei2006} found an expression of the electroviscous force $F_{ev,\parallel}$ exerted on a spherical particle translating parallel and close ($d \ll a$) to a solid plane with no-slip BC and with Zeta potential $\zeta_w$ \cite{tabatabaei2006,tabatabaei2010}:

\begin{align}
F_{ev,\parallel} =& C\dfrac{a^2}{d^2}\bigg[ \dfrac{8}{25}\big( G_+\left(7G_p + 2G_w \right) + H_+\left(7H_p + 2H_w \right)\big)\left(v+a\omega\right)  + && \nonumber \\ 
&-\dfrac{8}{5}\big( G_-\left(\alpha_1G_p + \alpha_2G_w \right) + H_-\left(\alpha_1H_p + \alpha_2H_w \right)\big)\left(v-a\omega\right) \bigg] \,, &&
\label{form:ev-drag_parallel}
\end{align}

\noindent where $G_\pm=G_p \pm G_w$, $H_\pm=H_p \pm H_w$, $\alpha_1=10.80625$, $\alpha_2=4.94467$, and $v$ and $\omega$ are the particle translational and angular velocities. Constants $G_w$ and $H_w$ are defined for the wall in the same way as $G_p$ and $H_p$ are defined for the particle in Eq. \ref{form:ev-drag_bulk_defGH1} and \ref{form:ev-drag_bulk_defGH2}.

%---

To the best of our knowledge, a theoretical prediction of electroviscous drag in the vicinity of a free interface is still missing. The development of a suitable model is far from being straightforward and is beyond the scope of the present paper. Accordingly, here we simply make a comparison between the expected order of magnitudes and trends of hydrodynamic and electroviscous drags.

It is first possible to evaluate Eq. \ref{form:ev-drag_bulk2} for the colloids and the solution considered in the present paper. In order to do this, we measured the particles Zeta potential as a function of molarity in the molarity range $\rho\in\left[10^{-6},10^{-3}\right]$ mol$\cdot$l\textsuperscript{-1} explored in the experimental campaign \cite{villa2020multistable}, obtaining $0\leq\zeta_p\leq-12.5$ mV. Accordingly, the maximum value assumed by $\xi_{ev}$ is $\xi_{ev} \simeq 3\cdot 10^{-12}$ kg$\cdot$s\textsuperscript{-1}, more than $4$ orders of magnitudes lower than the bulk hydrodynamic drag $6\pi\eta a \simeq 7\cdot 10^{-8}$ kg$\cdot$s\textsuperscript{-1}.
Electroviscous effect in the bulk is therefore expected to be negligible.

Close to the interface the dimensioned pre-factors of both hydrodynamic and electroviscous drags have the same orders of magnitude of the corresponding bulk cases. 
The additional element that has to be considered is the distance dependence of the drag. Considering small values of $h=d/a$, where lubrication approximation can be applied, hydrodynamic drag scales as $h^{-1}$. On the other side, in Eq. \ref{form:ev-drag_parallel} electroviscous drag in the wall vicinity scales as $h^{-2}$. The first one is therefore expected to increase faster than the latter as $h$ lowers: the ratio between the two drags is expected to be proportional to $h^{-1}$. However, the lowest values of $h$ experimentally explored in the present work are of the order of $0.01$, thus corresponding to a gain of $2$ orders of magnitudes in the ratio between electroviscous and hydrodynamic drags.  Although large, such a gain is still not enough to compensate for the low value of the ratio between the corresponding parameters $C$ and $6\pi\eta a$. These qualitative arguments allow us to safely exclude electrokinetic effects from the study of the particles dynamics treated in the present work.

\section{Effect of Surface Active Species on Brownian Dynamics: Modelling}
\label{appendix_model}
In the present appendix the effect on the Brownian dynamics normal to an air-water interface of a small surface concentration of surface active species $c_0$ is theoretically addressed.

\subsection{Frequency-Dependent Hydrodynamic Mobility}

Following Maali \textit{et al.}~\cite{maali2017viscoelastic}, we assume that the concentration of surface impurities $c$ is governed by an advection-diffusion equation of the form 

\begin{equation}
\frac{\partial c}{\partial t} + \boldsymbol{\nabla} \cdot \left( \bm{v}_\mathrm{S} c \right)
	= D_c \boldsymbol{\nabla}^2 c \, , \label{EqDiffAdv}
\end{equation}

\noindent with~$D_c$ denoting the diffusion coefficient of surface impurities and $\bm{v}_\mathrm{S} := \bm{v} (r, z=0)$ in the system of axisymmetric cylindrical coordinates $(r,z)$.
Equation~\eqref{EqDiffAdv} is subject to the boundary conditions of $v_z (z=0) = 0$ and $\eta \partial_z v_r (z=0) = \partial_r \Pi$, where $\Pi$ is the pressure associated with the presence of surface impurities.
In the present work, we employ the usual lubrication approximation and assume a quadratic evolution of the height of the fluid film between the colloid and the interface of the form $h(r) = d + r^2 / \left( 2a \right)$.
Here, $d$ is the distance to the air-water interface, measured from the bottom of the sphere of radius~$a$.
In addition, by restricting ourselves to the situation in which $\left| \bm{v}_\mathrm{S} \cdot \boldsymbol{\nabla} c \right| \ll \left| c \boldsymbol{\nabla} \cdot \bm{v}_\mathrm{S} \right| \simeq \left| c_0 \boldsymbol{\nabla} \cdot \bm{v}_\mathrm{S} \right|$, with $c_0$ denoting the equilibrium concentration (see Maali \textit{et al.}~\cite{maali2017viscoelastic} for more details), Eq.~\eqref{EqDiffAdv} can then be expressed in temporal Fourier space as

\begin{equation}
	i\omega c - \frac{\Gamma}{4\eta} \frac{1}{r} \frac{\partial}{\partial r} \left( rh \frac{\partial c}{\partial r} \right) - 
	\frac{3c_0 V\left(\omega\right) d}{2h^2} = 0 \, , \label{PDE}
\end{equation}
wherein $V$ is the frequency-dependent translational velocity of the particle normal to the interface.
Here, we have used the abbreviation $\Gamma = c_0 \kBT$.
A closed analytical solution of Eq.~\ref{PDE} is rather delicate and far from being trivial.
Therefore, to be able to make analytical progress, we attempt to obtain an approximate solution.

In the quasi-steady limit of vanishing frequency, the solution of Eq.~\eqref{PDE} that satisfies the underlying boundary conditions is given by $c(r, \omega=0) = K/h$, where $K = 3\eta VR/\left( k_\mathrm{B} T \right)$.
For arbitrary frequency, we assume as an Ansatz that an approximate solution of Eq.~\ref{PDE} can be searched for as a finite series of terms in inverse powers of~$h$ in the following form
% \textcolor{red}{[why we choose this form]}
\begin{equation}
	c(r, \omega) \simeq \sum_{n=1}^{N} \frac{a_n (\omega) + i b_n (\omega)}{h(r)^n} \, ,
	\label{c_form}
\end{equation}
where $a_i$ and~$b_i$ are frequency-dependent real quantities that are independent of~$r$.
Notably, Eq.~\ref{c_form} satisfies the regularity conditions $\frac{\partial c}{\partial r} = 0$ at $r=0$ and $c=0$ for $r \to \infty$.
In the following, we outline the derivation steps when taking only the first term in the series expansion describing the evolution of the concentration field. An analogous derivation approach can be followed to obtain more accurate solutions by taking a larger number of terms in the series expansion given by Eq.~\ref{c_form}.
%We have checked that the solution converges quickly by taking only a few terms.
%For the purpose of comparison with experimental results, however, restrict ourselves for simplicity to the leading term in this expansion is a .

Substituting the solution form given by Eq.~\ref{c_form} for $N=1$ and expanding the resulting expression around $r=0$ up to $\mathcal{O} \left( r^2 \right)$ yields 
\begin{equation}
	\left( \frac{\Gamma}{2\eta a} + i\omega \right) \left( a_1 + ib_1 \right)-\frac{3c_0 V}{2} = 0 \, . \label{equate1}
\end{equation}
By equating the real and imaginary parts on the left-hand side of Eq.~\ref{equate1} to zero and solving the resulting linear system of equations for the unknowns $a_1$ and $b_1$, we readily obtain
\begin{subequations}
	\begin{align}
	a_1(\omega) &= \frac{3\eta R c_0 V \Gamma}{\Gamma^2 + \left( 2\eta a\omega \right)^2} \, , \\
	b_1(\omega) &= -\frac{6\eta^2 a^2 c_0 V \omega}{\Gamma^2 + \left( 2\eta a\omega \right)^2} \, .
	\end{align}
\end{subequations}
Then, the solution of Eq.~\eqref{PDE} for the concentration field of surface active species at the interface can conveniently be approximated in Fourier space as
\begin{equation}
	c(r,\omega) = \frac{3\eta a c_0 V}{\left( \Gamma + 2i\eta a \omega \right) h(r)}\, .	\label{c1_sol}
\end{equation}

Moreover, as outlined in Maali \textit{et al.}~\cite{maali2017viscoelastic}, the hydrodynamic pressure is governed by the ordinary differential equation:
\begin{equation}
	\frac{\partial p}{\partial r} = 
	\frac{3}{2h} \left( \frac{\eta V r}{h^2} - \frac{\partial \Pi}{\partial r} \right) \, , \label{eq_pressure}
\end{equation}
with $\Pi$ denoting the surface pressure resulting from the presence of surface active species, resulting in $\Pi = c\kBT$ for an ideal gas.
Solving Eq.~\ref{eq_pressure} upon substitution of the solution for the concentration given by Eq.~\ref{c1_sol}, the pressure field up to a constant is obtained as
\begin{equation}
	p(r,\omega) = -\frac{3\eta a V \left( 2\Gamma + i\eta a \omega \right)}{2h(r)^2 \left( \Gamma + 2i\eta a \omega \right)} \, .
\end{equation}

Finally, the hydrodynamic force exerted on the translating particle is obtained upon surface integration as
\begin{equation}
	F^\mathrm{H}\left(\omega\right) = 2\pi \int_0^\infty p(r,\omega) \, r \, \Intd r  \, .
\end{equation}

Defining the drag coefficient as $\gamma := -F^\mathrm{H}/V$, we obtain
\begin{equation}
	\gamma (\omega) = \frac{3\pi \eta a^2 \left( 16\omega_c + i\omega \right)}{2d \left( 4\omega_c + i\omega \right)} \, , 
	\label{draglubrication}
\end{equation}
where we have defined $\omega_c = \Gamma / \left( 8\eta a \right) $.
In particular, we recover the drag coefficients near a no-slip interface $\gamma^\mathrm{NS} = 6\pi\eta a^2/d$ in the limit $\omega \to 0$ and near a free interface with full slip $\gamma^\mathrm{FS} = 3\pi\eta a^2/\left( 2d \right)$ in the limit $\omega \to \infty$.
These results are valid in the lubrication limit such that $d \ll a$.

An analogous derivation approach can be followed to obtain more accurate solutions by taking a larger number of terms in the series expansion given by Eq.~\ref{c_form}.
We have checked that the solution converges quickly by taking only a few terms.
For the purpose of comparison with experimental results, we restrict ourselves for simplicity to the leading term in this expansion.

\subsection{Generalized Langevin Dynamics}

The equation governing the dynamics of a Brownian particle in the presence of a confining well potential can be described using a generalized Langevin equation of the form~\cite{mankin2011generalized}
\begin{equation}
	m\ddot{x} + \int_0^t \gamma (t-t') \dot{x} (t') \, \Intd t'
	+ m\omega_0^2 x = F(t) \, , \label{GLE}
\end{equation}
with $\dot{x} \equiv \Intd x / \Intd t$, $x(t)$ is the displacement of the particle of mass~$m$, $\gamma (t)$ is the friction memory kernel, $\omega_0$ is the frequency of the harmonic driving force, and $F(t)$ is the stochastic random force that models the effect of the background noise.
The latter is assumed to be Gaussian distribute delta correlated with zero mean. 
Following Kubo's approach, Eq.~\ref{GLE} can be written in Fourier space as
\begin{equation}
	\Bigg( im\omega \left( 1- \frac{\omega_0^2}{\omega^2} \right) + \gamma[\omega]  \Bigg) V(\omega) = F(\omega) \, ,
\end{equation}
where we have used the properties of the Fourier transform of a derivative $V(\omega) = i\omega x(\omega)$ and denoted by $\gamma[\omega]$ the one-sided Fourier transform; also sometimes called Fourier-Laplace transform of~$\gamma(t)$.
It follows from the fluctuation-dissipation theorem~\cite{kubo66} that the velocity auto-correlation function (VACF) can be obtained from an inverse Fourier transform of the form~\cite{kubo85}
\begin{equation}
	\phi (t) = \frac{\kBT}{2\pi} \int_{-\infty}^{\infty} \left( \frac{i\omega}{m \left( \omega_0^2 - \omega^2 \right) + i\omega \gamma(\omega)} + c.c. \right) e^{i\omega t} \, \Intd \omega \, , %\notag
    \label{VACF}
\end{equation}
with $c.c.$ denoting the complex conjugate.
Here, we have assumed that $\gamma(t) = 0$ for $t<0$ so that $\gamma[\omega] \equiv \gamma(\omega)$.

In the overdamped regime, Eq.~\ref{VACF} takes the following particularly simple form~\cite{daddi16,epje_review}
\begin{equation}
	\phi (t) = \frac{\kBT}{2\pi} \int_{-\infty}^{\infty} \left( \frac{i\omega}{m \omega_0^2 + i\omega \gamma(\omega)} + c.c. \right) e^{i\omega t} \, \Intd \omega \, . 
\end{equation}

Finally, the particle mean-square displacement (MSD) can be obtained from the VACF as

\begin{equation}
	\mathrm{MSD} := \langle \left[x(t+\tau)-x(t)\right]^2 \rangle_t = 2 \int_0^\tau \left( \tau-t \right) \phi(t) \, \Intd t \, .
\end{equation}

For particle translational motion normal to a surface active species-covered interface, the drag coefficient in Fourier space is given in the lubrication limit by Eq.~\ref{draglubrication}.
The frequency-dependent hydrodynamic mobility is obtained in Fourier space as $\mu^S(\omega) = 1/\gamma(\omega)$.
Assuming that the evolution of the drag between the no-slip and free-slip limits occurs at the same time scale, the frequency-dependent hydrodynamic mobility in Eq.~\ref{draglubrication} can be generalized to
\begin{equation}
	\mu^S(\omega) = \frac{\eta a \omega \muFS + 2i \Gamma \muNS}{\eta a \omega + 2i \Gamma} \, ,
	\label{mu_general}
\end{equation}
with $\muNS$ and $\muFS$ denoting the mobilities near a no-slip and free-slip interface, respectively, for which exact analytical expressions are available~\cite{brenner1961slow}.

Substituting Eq.~\ref{mu_general} into Eq.~\ref{VACF} yields after integration an analytical expression for the MSD.
It can be cast in the form
\begin{equation}
	\mathrm{MSD} = Z
	\left( 2 \Lambda + 4 \Gamma - X_+ e^{-\frac{Y_+ \tau}{2\eta a}} - X_- e^{-\frac{Y_- \tau}{2\eta a}} \right) ,
	\label{eq:finalmsd}
\end{equation}
where we have defined $\Lambda = \eta a m \omega_0^2 \muFS$ and 
\begin{equation}
Z = \frac{m \kBT}{W} \left( \frac{8 \Gamma \eta a \omega_0 \muNS }{Y_+ Y_-} \right)^2 , 
\end{equation}
where

\begin{equation}
    W = \sqrt{\Lambda^2 - 4 \Gamma\Lambda \left( 1 - \frac{2\muNS}{\muFS} \right) + 4\Gamma^2}.
\end{equation}

In addition, $X_\pm = \Lambda + 2\Gamma \pm W$ and $Y_\pm = \mp \Lambda \pm 2\Gamma + W$.

In particular, for $\omega_0 = 0$, we obtain the solution in the absence of the confining potential.
Specifically,
\begin{equation}
	\mathrm{MSD} = \frac{\eta a}{c_0} \left( \muFS - \muNS \right) \left( 1-e^{-16\omega_c \tau} \right) 
	+ 2 \kBT \muNS \tau \, .  %\notag
	\label{eq:MSDfinalFree}
\end{equation}

For $c_0 = 0$, we obtain
\begin{equation}
	\mathrm{MSD} = \frac{2 \kBT}{m \omega_0^2} \left( 1 - e^{-m \omega_0^2 \muFS \tau} \right)\, , %\notag
\end{equation}
corresponding to the expected MSD for the confined motion orthogonal to a free interface in the overdamped regime. % (equation \ref{eq_msd_harmonic_potential}).

\begin{figure}
    \centering
    \includegraphics[width=8.6cm]{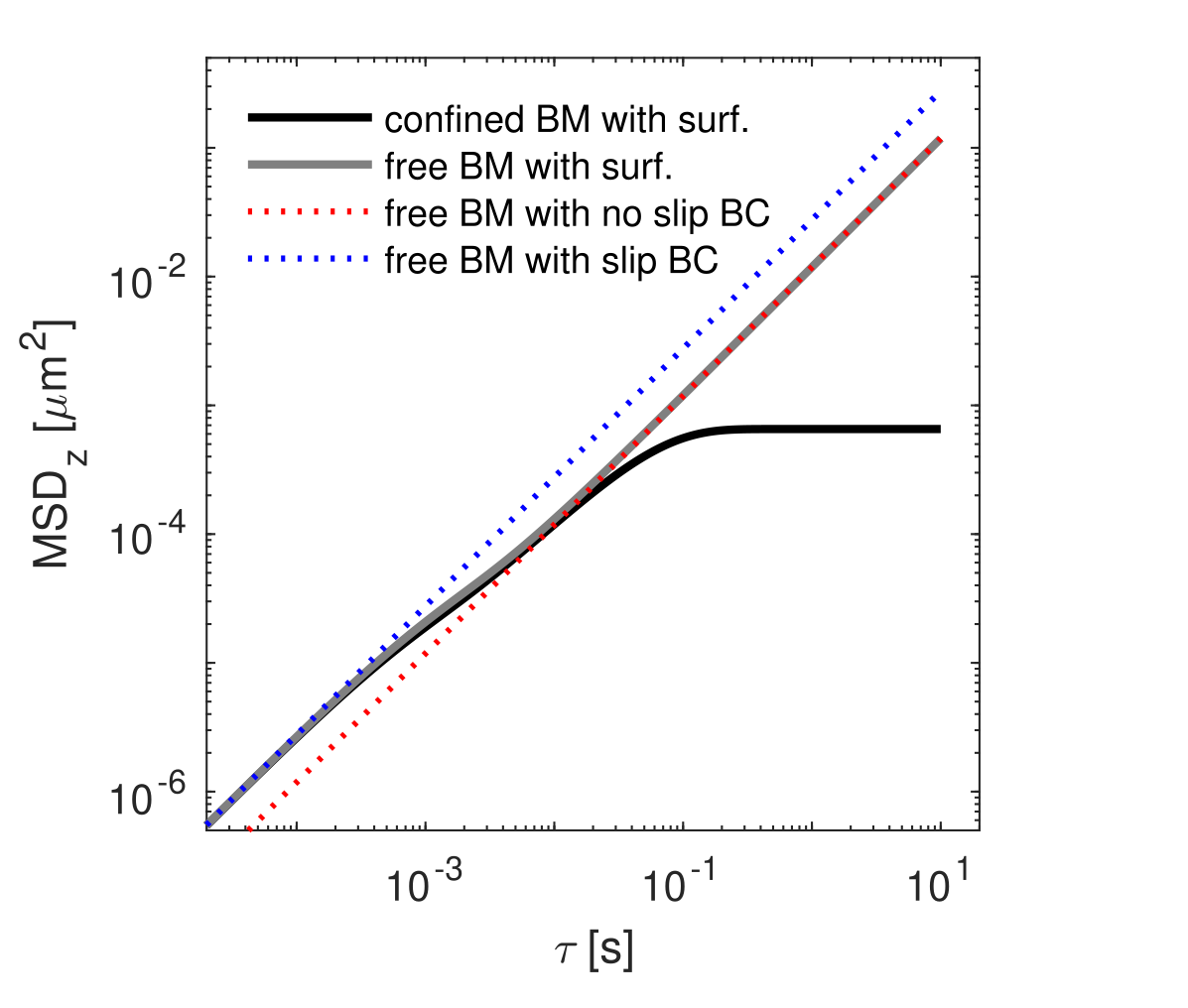}
    \caption{MSDs for a sphere moving orthogonal to a surface active species-covered interface according to equations \ref{eq:finalmsd} (black) and \ref{eq:MSDfinalFree} (gray). Dotted lines represent expected MSDs for free Brownian dynamics orthogonal to a full slip (blue) and to a no-slip (red) interface.}
    \label{fig:appe}
\end{figure}

The general MSD for a bead translating orthogonal to a surface active species-covered interface and confined in a potential well reported in eq. \ref{eq:finalmsd} is a sum of two exponentials with characteristic times $1/\omega_\pm$, where frequencies $\omega_\pm$ are defined as $\omega_\pm := Y_\pm/2\eta a$. 
In the overdamped regime and limiting to the experimental case where $\omega_c\gg \omega_\mu := (2m\mu)^{-1}$, $\omega_\pm$ can be easily related to the typical system timescales. Indeed, under such limits it can be shown that $W\sim 2\Gamma$ and thus:
\begin{subequations}
    \begin{align}
        \omega_- \sim \frac{\omega_0^2}{2\omega_\mu}
    \end{align}
    \begin{align}
        \omega_+ \sim 16\omega_c
    \end{align}
\end{subequations}

In equation \ref{eq:finalmsd}, therefore, the first exponential represents the transition from full slip to no-slip BC resulting from the presence of surface active species, while the second one describes the DLVO confinement of the sphere Brownian dynamics.
In Figure \ref{fig:appe} are reported examples of equations \ref{eq:finalmsd} (black) and \ref{eq:MSDfinalFree} (gray) evaluated for the same experimental parameters corresponding to the data reported in Figure \ref{fig:phantom}. In both cases the initial transition from full slip (blue) to no-slip (red) can be observed between $10^{-3}$ s and $10^{-2}$ s.

\bibliographystyle{unsrt} % We choose the "plain" reference style

%---------------------------------------------------------------------------------
%                                   BIBLIOGRAPHY
%---------------------------------------------------------------------------------
%\bibliography{biblio}
\providecommand{\noopsort}[1]{}\providecommand{\singleletter}[1]{#1}%

%---------------------------------------------------------------------------------
%---------------------------------------------------------------------------------

\end{document}